\documentclass{pnastwo}

\def\be{\begin{equation}}
\def\ee{\end{equation}}

\def\bi{\begin{itemize}}
\def\ei{\end{itemize}}
\def\bn{\begin{enumerate}}
\def\en{\end{enumerate}}
\def\bea{\begin{eqnarray}}
\def\eea{\end{eqnarray}}
\newcommand{\bpm}{\begin{pmatrix}}
\newcommand{\epm}{\end{pmatrix}}

\def\ba{\begin{array}}
\def\ea{\end{array}}
\def\bd{\begin{displaymath}}
\def\ed{\end{displaymath}}

\renewcommand{\imath}{\hspace{1pt}\mathrm{i}\hspace{1pt}}

\url{www.pnas.org/cgi/doi/10.1073/pnas.0709640104}
\copyrightyear{2008}
\issuedate{Issue Date}
\volume{Volume}
\issuenumber{Issue Number}

\begin{document}

\title{Are the surface Fermi arcs in Dirac semimetals topologically protected?}

\author{Mehdi Kargarian\affil{1}{Department of Physics, The Ohio State University, Columbus, OH 43210},
Mohit Randeria\affil{1}{},
\and
Yuan-Ming Lu\affil{1}{}}

\contributor{Submitted to Proceedings of the National Academy of Sciences
of the United States of America}

\significancetext{In recent years there has been a surge of interest in quantum materials with topologically protected properties that are robust against the effects of disorder and other perturbations. An interesting example is the newly discovered three dimensional analog of graphene, called Dirac semimetals. These were expected to have highly unusual conducting states on their surfaces. We show here that the surface states of Dirac semimetals do not exhibit the expected topological robustness and, quite generally, get deformed into states that are of a rather different character. Our theoretical results not only have conceptual importance in the field of topological quantum materials but also make clear predictions that can be tested in several experiments.
}

\maketitle

\begin{article}
\begin{abstract}
{
Motivated by recent experiments probing anomalous surface states of
Dirac semimetals (DSMs) Na$_3$Bi and Cd$_3$As$_2$, we raise the question posed
in the title. We find that, in marked contrast to Weyl semimetals, the gapless surface states
of DSMs are not topologically protected in general, except on time-reversal-invariant planes of surface Brillouin zone.
We first demonstrate this in a minimal $4$-band model with a pair of Dirac nodes at ${\bf k}=(0,0,\pm Q)$, where
gapless states on the side surfaces are protected only near $k_z=0$.
We then validate our conclusions about the absence of a topological invariant protecting
double Fermi arcs in DSMs using a K-theory analysis for space groups of Na$_3$Bi and Cd$_3$As$_2$.
Generically, the arcs deform into
a Fermi pocket, similar to {{the surface states of a topological insulator (TI), and this can merge into the}}
projection of bulk Dirac Fermi surfaces as the chemical potential is varied.
We make sharp predictions for the doping-dependence of the surface states of a DSM
that can be tested by ARPES and quantum oscillation experiments.
}
\end{abstract}

\keywords{Dirac semimetals | Weyl semimetals | Topological insulators | Fermi arcs}


\dropcap{F}ollowing the theoretical prediction and experimental discovery of topological insulators~\cite{Hasan:rmp10,Qi:rmp11} in the past decade,
there has been an explosion of interest in understanding the role of topology in various quantum states of matter.
An important set of questions concerns the topological properties of {\it gapless} Fermi
systems~\cite{volovik:book,Horava2005,vishwanath:prb11,burkov:prb11,Zhao2013,Matsuura2013}.
Of particular interest are three-dimensional (3D) semimetals where the bulk electronic dispersion exhibits point nodes at the Fermi level
and the low energy physics is effectively described by  Weyl or Dirac Hamiltonian~\cite{vafek:annal14}.

A striking feature of 3D Weyl semimetals (WSM), which necessarily break either time-reversal or inversion symmetry,
is the existence~\cite{vishwanath:prb11} of topologically protected surface ``Fermi arcs''. Here the Fermi contour
in the surface Brillouin zone breaks up into disconnected pieces, which connect the projection
of two Weyl nodes with opposite chirality. The Fermi arcs have been recently observed in angle-resolved photoemission spectroscopy (ARPES)
studies of non-centrosymmetric TaAs~\cite{Xu:Weyl_science15,weng:prx15,huang:nc15,lv:prx15,Lv:nature15,Xu:nature15}.

Here we focus on another outstanding example, the Dirac semimetal (DSM), which is the 3D analogue of graphene. In the presence of both
time-reversal and inversion symmetries, the electronic excitations near each node are described by a four-component Dirac fermion, and
a dispersion that is linear in all directions in ${\bf k}$-space~\cite{zaheer:prl12,wang:prb12,wang:prb13}.
ARPES has clearly observed linearly-dispersing bands near Dirac nodes in two DSM materials Na$_{3}$Bi~\cite{Li:science14,cava:Na3Bi} and
Cd$_{3}$As$_{2}$~\cite{Yi:Sciencereport14,Neupane:ncomm14,Jeon:nmat14}.
The signatures of Fermi arcs by ARPES in Na$_{3}$Bi~\cite{xu:science15} and by their peculiar quantum oscillations~\cite{Potter2014}
in Cd$_{3}$As$_{2}$~\cite{analytis:arXiv1505.02817} have recently been reported.
{Although DSM can in principle appear in a system with spin rotational symmetry~\cite{Morimoto2014},
here we focus on DSMs in spin-orbit-coupled systems which are more closely related to material realizations.}

We can understand the Dirac fermions as two degenerate Weyl fermions with opposite chirality, where crystal symmetries
forbid the two Weyl nodes from hybridizing and opening up a gap at each Dirac point~\cite{zaheer:prl12,yang:ncomm2014}. Given this picture of the bulk, it is natural to expect the surface states in a DSM~\cite{wang:prb12,wang:prb13} as two copies of the chiral Fermi arc on WSM surface: i.e. the ``double Fermi arcs'' shown schematically in Fig.~\ref{fig:prl_fig}(a).

In this paper, we address the important question of whether the the double Fermi arcs on the surface of Dirac semimetals topologically protected? If yes, what is the associated topological invariant? The Fermi arcs in WSMs are robustly protected for the following reason: each plane that lies between a pair of Weyl nodes in momentum space, perpendicular to the separation between them, has an integer Chern number~\cite{vishwanath:prb11,Yang2011,nagaosa:prl14} associated with quantum Hall effect. The chiral edge modes of these momentum-space Chern insulators give rise to robust surface Fermi arcs, which end at the projection of bulk Weyl nodes. Unlike the WSM, the bulk-boundary
correspondence in topological phase~\cite{Matsuura2013} provides no obvious answer for DSMs, since the stability of bulk Dirac nodes
require crystal rotation symmetries~\cite{wang:prb12,wang:prb13} which are explicitly broken on open side surfaces.
It was claimed~\cite{Potter2014} though, that the surface Fermi arcs in DSMs are perturbatively stable against a weak symmetry-breaking surface potential, and a strong surface potential can destroy them.

Here we provide a surprising answer to the question posed in the title. We show that the double Fermi arcs on Dirac semimetal surface
are \emph{not} topologically protected, and can be continuously deformed into a closed Fermi contour
without any symmetry breaking or bulk phase transition; see Fig.~\ref{fig:prl_fig}.
{{The resulting Fermi contours do have topological character, even though they are
not as exotic as the Fermi arcs in WSMs. We show that the surface states of a DSM with
an odd (even) number of Dirac node pairs are analogous to the surface states of a 3D strong TI (weak TI or trivial insulator).
}}

To make our results physically transparent, we first focus on the 4-band minimal model for
DSMs~\cite{wang:prb12,wang:prb13,chiu:conference} and present simple arguments for lack of topological protection for surface Fermi arcs.
We then show numerical results for ARPES spectral functions for the 4-band model which demonstrates how a bulk perturbation that
respects all symmetries can destroy the Fermi arcs. Next, we go beyond the simple low-energy model and directly address the two DSM materials of experimental interest. We use K-theory~\cite{Horava2005,Kitaev2009,Wen2012,Morimoto2013, sato:prb14}
to classify the topological stability of surface states in the presence of time reversal, charge conservation and space group symmetries in
Dirac semimetals Cd$_3$As$_2$ and Na$_3$Bi. This classification indicates that topological protection for gapless surface states on
side surfaces (that do not intersect the $\hat{z}$-axis) of Dirac semimetals exists only in the high-symmetry plane $k_z=0$.
We conclude by examining the experimental implications of our results. We discuss how to understand the existing
ARPES and quantum oscillation measurements in light of our results on the lack of topological protection for Fermi arcs,
and we make sharp predictions for testing our conclusions.

\section{Minimal model}
We begin with a simple 4-band model for a Dirac semimetal with 4-fold rotational symmetry along the $\hat z$-axis defined by the Hamiltonian:
\bea \label{H4}
H(\textbf{k})&=&\varepsilon^0_\textbf{k}\!+\!\Big[t(\cos k_x\!+\!\cos k_y\!-\!2)\!+\!t_{z}(\cos k_z\!-\!\cos Q)\Big]\tau_{z}\nonumber \\
&&+\lambda \sin k_x\sigma_{x}\tau_{x}+\lambda \sin k_y\sigma_{y}\tau_{x}.
\eea
The Pauli matrices $\vec\sigma$ act on spin and $\vec\tau$ in orbital space,
$t$ and $t_z$ are hopping amplitudes and $\lambda$ is spin-orbit coupling (SOC).
Ignoring $\varepsilon_\textbf{k}^{0}$ for the moment, there is a gap everywhere in the Brillouin zone (BZ) except
at two points ${\bf k}=(0,0,\pm Q)$ where the low energy spectrum is described by linearly-dispersing Dirac fermions.

We can always add to eq.~(\ref{H4}) an $\varepsilon^0_\textbf{k}$ that preserves all the symmetries and vanishes at the Dirac nodes,
e.g., $\varepsilon^0_{\textbf{k}}=t_{1}(\cos k_{z}\!-\!\cos Q)\!+\!t_{2}(\cos k_{x}\!+\!\cos k_{y}\!-\!2)$
or $\varepsilon^0_{\textbf{k}}=t_{1}(\cos k_{z}\!-\!\cos Q)(\cos k_{x}\!+\!\cos k_{y})$. While this term does not
qualitatively change the Dirac  spectrum in the bulk, it gives rise to a curvature of the
Fermi arc in the surface BZ, as discussed below.

The symmetries of $H(\textbf{k})$ and their representations in terms of spin and orbital operators
are as follows. Time-reversal is implemented by the antiunitary operator $\Theta=\imath\sigma_{y}\cdot\mathcal{K}$, where
$\mathcal{K}$ is complex conjugation and $\Theta^2 = -1$.
Inversion $I$ corresponds to $U_{I}=\tau_{z}$, with the two orbitals having even and odd parity.
Two-fold rotation $C_{2,x(y)}$ about the $\hat{x}$ (or $\hat{y}$) axis is implemented by $U_{c_{2,x(y)}}=\imath\sigma_{x(y)}$,
and $n$-fold rotation $C_{n,z}$ by $U_{c_{n,z}}=\exp\left({\imath\pi\sigma_{z}(1-2\tau_z)/n}\right)$,
where $n=4$ here.
$H(\textbf{k})$ also has mirror reflection symmetries with respect to the $j=x,y,z$-planes,
with $k_{j}\rightarrow -k_{j}$, which is implemented by $U_{R_{j}}=U_{c_{2,j}} U_{I}=\imath\sigma_j\tau_z$.

\section{Instability of surface Fermi arcs}
{Surfaces with a normal not parallel to $\hat z$-axis are expected to support gapless states with Fermi arcs, shown schematically in Fig. \ref{fig:prl_fig}(a) for side surfaces}.
We address the lack of topological protection for the Fermi arcs of $H(\textbf{k})$ from three different points of view.
(i) First, we examine the stability of 1D (one-dimensional) edge states that exist on 2D slices of BZ at a fixed, generic value of $k_z$.
Here we define a generic $k_z$ to be $k_z \notin \{\pm Q, 0, \pi\}$, so that a fixed-$k_z$ plane will neither contain Dirac nodes, nor will
it have the additional symmetry arising from time-reversal invariance.
(ii) Next, we prove the existence of an extra mass term in the gapped Dirac Hamiltonian that describes the plane with a generic $k_z$.
(iii) Finally we show how adding fully symmetric, bulk perturbations to the 3D Hamiltonian $H(\textbf{k})$ can destroy the Fermi arcs,
without destroying the bulk Dirac nodes. The projection of these bulk nodes onto the surface does remain gapless, as do the 2D slices located at
$k_z=0$ as they have extra (time-reversal) symmetry.

Without loss of generality we consider a (100) surface. This surface preserves only the following symmetries: time reversal $\Theta$, mirror reflections $R_y,R_z$ and their combinations. Therefore a fixed-$k_z$ plane in a system with with a (100) surface only respects two symmetry operations: mirror reflection $U_{R_{y}}=\imath\sigma_y\tau_z$ w.r.t. $\hat y$-plane, and the combination $\Theta_{R_z}=\Theta U_{R_{z}}=\imath\sigma_{x}\tau_{z}\cdot\mathcal{K}$ of time reversal and mirror w.r.t. $\hat z$-plane. Note that $\Theta_{R_z}^2=1$.
Under these symmetries the Hamiltonian at fixed $k_z$, denoted by $\tilde{H}(k_x,k_y)$, transforms as
$\Theta_{R_z}^{-1}\tilde{H}(k_x,k_y)\Theta_{R_z}=\tilde{H}(-k_x,-k_y)$ and
$U_{R_{y}}^{-1} \tilde{H}(k_x,k_y) U_{R_{y}}=\tilde{H}(k_x,-k_y)$.

(i) First consider the surface BZ $(k_y,k_z)$ for a $(100)$ surface. If we look at a 1D slice at fixed-$k_z$ where we
might have expected gapless edge states described by
$H_{\rm edge}=\hbar v_{F}\sum_{k_{y}}\psi_{k_y}^{\dag}(k_{y}\mu_{z}-k_{F})\psi_{k_{y}}$.
Here $\psi^{T}_{k_y}=(\psi_{R,k_{y}}, \psi_{L,k_{y}})$ is a two-component spinor of right (R) and left (L) movers
and $\vec\mu$ are Pauli matrices in $(R,L)$-space. Note that the dispersion $v_F$ and the location of the Fermi arc $k_F$ are both,
in general, $k_z$-dependent.

The two symmetry operations on the edge states are represented by
$\Theta_{R_z}=\imath\mu_x\cdot\mathcal{K}$ and $U_{R_y}=\imath\mu_y$.
We now add perturbation $\hat{M}=m\sum_{k_{y}}\psi_{k_y}^{\dag}\mu_{y}\psi_{k_{y}}$,
which preserves both these surface symmetries: $[\Theta_{R_{z}},\mu_{y}]=[U_{R_{y}},\mu_{y}]=0$.
Clearly $\hat{M}$ is a mass term for $H_{\rm edge}$ that destroys the gapless edge states.

(ii) Next, we gain insight into the instability of surface states on a fixed-$k_z$ plane within
classification scheme~\cite{Kitaev2009,Wen2012,Morimoto2013} of topological insulators.
For a fixed, generic $k_z$, our system is a gapped 2D insulator described by a Dirac Hamiltonian
\bea
\tilde{H}_D=k_{x}\gamma_{x}+k_{y}\gamma_{y}+m\gamma_{0},
\eea
up to an additive constant term.
Here $\gamma_{x}=\sigma_{x}\tau_{x}$, $\gamma_{y}=\sigma_{y}\tau_{x}$ and $\gamma_{0}=\tau_{z}$ are Dirac matrices
in the minimal model. Let us ask if we can add another mass term $m'\gamma_{0}'$ to $\tilde{H}_D$
which preserves all the symmetries and anti-commutes with all the Dirac matrices $\gamma_{x,y,0}$.
If we can do this, the $m>0$ phase  can be continuously deformed into the $m<0$ phase without closing the gap at $m=0$,
and $\tilde{H}_D$ would be topologically trivial.
Such a term indeed exists: $\gamma_0^\prime=\sigma_z\tau_x$ commutes with $\Theta_{R_z}$ and $U_{R_y}$ and
anticommutes with $\gamma_{x,y,0}$. Therefore the surface states at generic $k_z$ are not topologically protected.

In contrast to generic $k_{z}$, the time-reversal-invariant planes $k_z=0$ and $k_z=\pi$ are gapped 2D insulators
with the well-known $\mathbb{Z}_2$ topological index associated with quantum spin Hall (QSH) effect~\cite{Hasan:rmp10,Qi:rmp11}.
In the DSM materials Cd$_3$As$_2$ and Na$_3$Bi, there is a nonsymmorphic glide reflection $g_c$ containing half-translation
along $\hat z$-axis, which serves as a mirror reflection for $k_z=0,\pi$ planes satisfying $(g_c)^2=-1(+1)$ for $k_z=0(\pi)$.
Generically such nonsymmorphic glide reflections guarantees that the $\mathbb{Z}_2$ index must be trivial for the $k_z=\pi$
plane~\cite{Varjas2015},  which will be elaborated below.
Therefore in these materials, only $k_z=0$ plane can support a nontrivial $\mathbb{Z}_2$ index.
A simple calculation~\cite{Fu:prb07} shows that the $k_z=0$ plane is indeed a nontrivial QSH insulator,
and gapless surface states at $k_z=0$ are topologically protected.

(iii) Finally we directly examine the gapless surface states of the 3D Hamiltonian (\ref{H4})
in a slab geometry with a surface at $x=0$, which is translationally invariant along $y$ and $z$.
We numerically calculate the spectral density $A(\textbf{k},\omega)=-(1/\pi)\rm Im G(\textbf{k},\omega\!+\imath0^+)$ associated with the
electron Green function $G(\textbf{k},\omega)$.
Here we measure $\omega$ with respect to the zero chemical potential, which coincides with the bulk Dirac nodes.
We show in Fig.~\ref{fig:prl_fig} the surface [panels (b,c,d)]
contributions to $A(\textbf{k},0)$ as a function of in-plane momentum $(k_{y},k_{z})$.
Note that ARPES probes precisely this spectral density multiplied by the Fermi-Dirac function.

In Fig.~\ref{fig:prl_fig}(b), we plot the result for $H(\textbf{k})$ of eq.~(\ref{H4})
and clearly see the two Fermi arcs extending from one Dirac node to the other.
In the absence of the $\varepsilon^0_\textbf{k}$ term of the form discussed above,
$H(\textbf{k})$ has an (unnatural) ``chiral symmetry" and the two Fermi arcs
become ``degenerate'' and collapse to $k_y = 0$ for $-Q\!<\!k_z\!<\!Q$~\cite{suppl}. 
With $\varepsilon^0_\textbf{k}$ in place, we get the arcs shown in Fig.~\ref{fig:prl_fig}(b).

Now we ask if there is a perturbation that has the full symmetry, does not shift the bulk nodes
and yet has a $\sigma_\alpha\otimes\tau_\beta$ structure that anticommutes with each term in
$H(\textbf{k})$ (other than the $\varepsilon^0_\textbf{k}$ term, of course). Such a perturbation
does exist and it must be of the form
\bea\label{H4:sym pertb}
\delta H_4({\bf k})=m^\prime(\cos k_{x}-\cos k_{y})\sin k_{z}\ \sigma_{z}\tau_{x}
\eea
This perturbation destroys the Fermi arcs near the projections of the Dirac nodes at $k_z = \pm Q$
on to the surface BZ as shown in Fig. \ref{fig:prl_fig}. We clearly see this in panel (c) where $\delta H_4({\bf k})$ with
$m'=0.4t$ deforms the Fermi arcs into a closed Fermi pocket. An increase in the strength of the perturbation to $m'=0.8t$
shrinks the arcs to even smaller pockets as shown in Fig. \ref{fig:prl_fig}(d). While the surface Fermi arcs are
progressively destroyed, the bulk Dirac nodes (and bulk states near $E_F=0$) are unaffected by these
perturbations. We note that this perturbation of ${\cal O}(k^3)$ is higher order than that
retained in usual ${\bf k}\cdot{\bf p}$ perturbation theory.

The minimal 4-band model~\cite{chiu:conference} is equivalent, up to a unitary transformation~\cite{suppl}, to the well-known
effective ${\bf k}\cdot{\bf p}$ Hamiltonian around the $\Gamma$ point for DSM materials~\cite{wang:prb12,wang:prb13}.
Despite its simplicity, this model with $n=4$-fold rotational symmetry about the $\hat{z}$-axis
captures the band inversion near the $\Gamma$ point of BZ in Cd$_3$As$_2$.
A similar model can also be adopted for the case of Na$_3$Bi with $n=6$-fold rotational symmetry~\cite{suppl}.

\section{K-theory}
Using K-theory we next show that the surface states and associated Fermi arcs
in Cd$_3$As$_2$ and Na$_3$Bi are not topologically protected.
This mathematical framework has the great  advantage of focusing
only on the space group, $U(1)$ charge conservation and time reversal symmetry of a material, without the need for a low-energy effective model
such as the one used in the analysis above. We thus obtain general and rigorous results that are directly applicable to real materials.

In the K-theory approach~\cite{Kitaev2009,Wen2012,Morimoto2013}, the classification of distinct gapped symmetric phases is reduced
to the following mathematical problem: what is the ``classifying space" of symmetry-allowed mass matrix for a generic Dirac Hamiltonian
preserving certain symmetries? Different symmetric phases correspond to disconnected pieces of the classifying space, which cannot
be continuously connected to each other without closing the bulk energy gap. Mathematically the group structure formed by these different
phases is given by the zeroth homotopy group $\pi_0(S)$ of classifying space $S$.
Here we only sketch the idea behind the calculations; detailed analysis are given in Supplementary Materials~\cite{suppl}.

{\bf Na}$_{3}${\bf Bi}: The hexagonal DSM material Na$_{3}$Bi has
a unit cell is defined by $\textbf{a}=a(1,0,0)$, $\textbf{b}=a(-1/2,\sqrt{3}/2,0)$ and $\textbf{c}=c(0,0,1)$.
Its space group is centrosymmetric P6$_{3}/$mmc (No.~194)~\cite{wang:prb12},
with its symmetry operations are generated by
6-ford screw $s_6$: $(x,y,z)\rightarrow (x-y,x,z+1/2)$,
glide reflection $g_c$: $(x,y,z)\rightarrow (y,x,z+1/2)$,
inversion $I$: $(x,y,z)\rightarrow (-x,-y,-z)$,
and Bravais lattice translations $T_{a,b,c}$.

The two Dirac nodes~\cite{wang:prb12} in Na$_{3}$Bi are located
at $\textbf{Q}_{\pm}=\pm Q_{0}(2\pi/c)(0,0,1)$. We consider two different surfaces which can potentially
contain nontrivial Fermi arcs. (i) A (010) surface spanned by the $\textbf{a}$ and $\textbf{c}$ axes,
and (ii) a (110) surface spanned by $(\textbf{a}-\textbf{b})$ and $\textbf{c}$.
In each case we proceed as follows.
We begin by determining the reduced set of symmetries that are preserved in the presence of the surface.
By analyzing how these symmetries act on a general Bloch Hamiltonian, we determine those symmetries that
preserve a 2D plane in the BZ zone at a fixed value of $k_z \neq 0,\pi/c$. All such planes, except those
passing through the nodes at $\textbf{Q}_{\pm}$ are fully gapped.

The basic strategy is to exploit the classification of gapped insulators described by Dirac Hamiltonians defined
on these 2D planes embedded in the 3D Brillouin zone. We use the
K-theory classification~\cite{Kitaev2009,Wen2012,Morimoto2013} of
gapped free-fermion quantum phases by examining the combined
algebra for the Dirac $\gamma$-matrices and the symmetry operations
and determining whether the \emph{classifying space} for symmetry-allowed mass terms is connected or not.

Using this approach we conclude that a generic 2D plane in the BZ
at fixed $k_z$ is topologically trivial and does not need to support protected surface states.
The plane $k_z = \pi/c$ is also shown to be trivial, due to the presence of the
nonsymmoprhic glide reflection $g_c$. Only the $k_z=0$ plane is found to have a nontrivial $\mathbb{Z}_2$ QSH index.

{\bf Cd}$_3${\bf As}$_2$:
The space group of the tetragonal DSM Cd$_{3}$As$_{2}$ is centrosymmetric $\mathrm{I4_{1}/acd}$ (No.~142)~\cite{Cava:IC2014}.
In addition to Bravais lattice translations, its symmetry operations are generated by
4-fold screw rotation along $\hat{z}$-axis: $(x,y,z)\rightarrow (y,1/2-x,1/4+z)$,
glide reflection $g_{c}$: $(x,y,z)\rightarrow (x,-y,1/2+z)$,
and inversion $I$: $(x,y,z)\rightarrow (-x,1/2-y,1/4-z)$, where $(s_4)^4=(g_{c})^2=-T_{z}$.
The Bravais lattice is expanded by $T_{\pm1,\pm1,\pm1}\equiv (\pm 1,\pm 1,\pm 1)/2$ similar to BCC lattice.
The two Dirac nodes in Cd$_{3}$As$_{2}$ are located along the $\hat z$-axis.

The K-theory procedure is used to determine the topological nature of various 2D planes in the BZ at fixed $k_z$,
in a manner completely analogous to that described above, taking into account of the symmetries of Cd$_{3}$As$_{2}$.
The calculations are described in supplementary materials~\cite{suppl}, and the conclusions are identical to the ones given above.
Only the $k_z=0$ plane has a nontrivial $\mathbb{Z}_2$ QSH index, all other fixed $k_z$ planes are trivial.

\section{Nature of Surface States}
We next show that the closed Fermi pocket on a side surface is essentially the same as the surface
Dirac fermion of a 3D strong topological insulator (TI). Consider a weak perturbation that breaks
$C_n$ symmetry, but preserves time reversal and inversion. We require the strength of this perturbation is smaller than the bulk gap at the time reversal invariant momenta (TRIM) of the 3D BZ. Clearly this perturbation will gap out the Dirac nodes but will not affect of parity eigenvalues of filled bands at the TRIM. Now that the bulk is fully gapped band insulator with time reversal symmetry, its $\mathbb{Z}_2$ topological invariant is given by the number of odd-parity Kramers pairs in filled bands at TRIM as shown by Ref.~\cite{Fu:prb07}. Since the $k_z=0$ plane is a 2D QSH insulator and $k_z=\pi$ is a trivial 2D insulator, the gapped bulk must be in a 3D strong TI phase. In the process of turning on the perturbation, the surface states at $k_z=0$ remain gapless since time reversal is always preserved. This demonstrates the equivalence between the closed Fermi pocket for DSMs and surface Dirac fermions of strong TIs.

{An important conclusion from our analysis is that if the bulk band touching
occurs at an odd number of pairs of Dirac nodes not located at TRIM (such as in Cd$_3$As$_2$ and Na$_3$Bi),
then the surface states cannot be fully removed due to time reversal symmetry.
Let $N_p$ Dirac points be located in the top half of the bulk BZ
($0<k_z<\pi$) and their $N_p$ time-reversal counterparts in the bottom half BZ ($-\pi<k_z<0$).
Based on our arguments, this will lead to $N_p$ closed Fermi pockets on a generic surface, which are either centered around $k_z=0$ or $k_z=\pi$.
If there are an odd number of Fermi pockets centered around $k_z=0$, the gapped $k_z=0$ plane in 1st BZ must correspond to a
2D QSH insulator with an odd number of helical edge states. Therefore, if $N_p$ is odd, $k_z=0$ and  $k_z=\pi$ must have opposite
$Z_2$ topological indices, with one plane being a trivial 2D insulator and the other a QSH insulator.
Hence the surface states of of an odd-$N_p$ DSM cannot be fully gapped out, in complete analogy with a 3D strong TI.
On the other hand, if $N_p$ is even, both the $k_z=0$ and $k_z=\pi$ planes share the same $Z_2$ index i.e. they are both trivial 2D insulators or both QSH insulators. In the former case there are generally no surface states, while gapless surface states in the latter case can be gapped out by translation symmetry breaking perturbations in analogy to a 3D weak TI.
As a result, an ``odd DSM'' with an odd number of pairs of Dirac points must support time-reversal-protected gapless surface states,
in contrast to an ``even DSM'', whose surface states can either be fully gapped or need protection from crystal translation symmetry.}

\section{Experimental Implications}
\begin{figure*}[!htb]
\includegraphics[width=17cm]{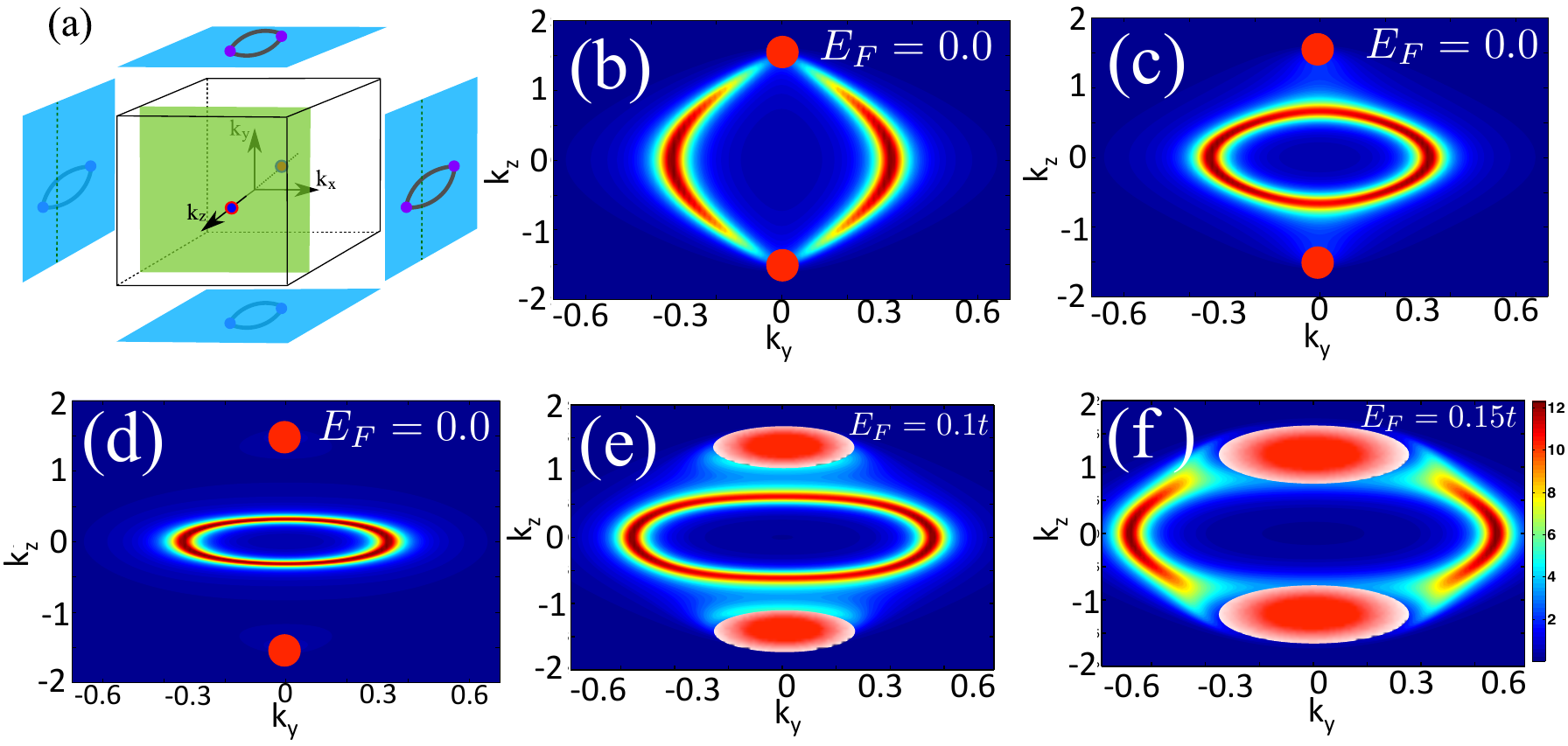}
\caption{(Color online) (a) Schematic {\bf k}-space picture of a Dirac semimetal showing Dirac nodes along $k_z$ axis in bulk Brillouin zone (BZ)
and possible double Fermi arcs on the surface BZ's, shown as blue squares. Note that surfaces perpendicular to $z$ axis have no arcs. A 2D slice of
the bulk BZ perpendicular to $k_z$ axis is shown as a green square, which projects to a green dashed line on a side surface.
(b) Surface spectral density of model in (\ref{H4}), which clearly shows the existence of double Fermi arcs on (100) surface.
(c,d) Continuous deformation of double Fermi arcs on the (100) surface by adding the perturbation $\delta H_{4}(\textbf{k})$ to (\ref{H4}).
(c) shows the effect at $m'=-0.4t$, while (d) corresponds to $m'=-0.8t$, showing that the Fermi arcs are progressively destroyed by increasing
strength of perturbation. The red solid circles in (b-d) correspond to the projection of bulk nodes and we set $E_{F}=0$ to line up the Fermi level with bulk Dirac nodes.
(e,f) correspond to electron doped systems with $m'=-0.8t$ (case in d) by raising the Fermi energy to $E_{F}=0.1t$ and $E_{F}=0.15t$, respectively.
The large red blobs in (e,f) mark the projection of bulk states onto the surface BZ.  The Fermi contour of the surface states
is disconnect from the blobs in panel (e), while it merges into the bulk states in panel (f).
} \label{fig:prl_fig}
\end{figure*}

We have shown above that, in general, the surface states of Dirac semimetals Cd$_3$As$_2$ and Na$_3$Bi have
a closed Fermi surface (FIG. \ref{fig:prl_fig} (c)), in contrast to the double Fermi arcs that one might naively expect (FIG. \ref{fig:prl_fig} (b)).
We now address the important question of how one can experimentally distinguish between these two types of surface states.

Assume, first, for simplicity that the chemical potential is at the Dirac point;
(we discuss below the case when it is not).
The qualitative differences between the two outcomes are then as follows.
The double Fermi arc has a singular change in slope when two arcs meet at
the projection of a Dirac node on to the surface BZ. In contrast, a Fermi contour has a smooth curvature everywhere in the surface BZ
and it does not pass through the two points that are the projections of the bulk Dirac nodes.
In addition, the wave-function associated with the surface states merges with the bulk at the tip of the arc, namely the Dirac node;
in contrast the states associated with a regular Fermi contour are exponentially localized at the surface.
All of these features could, in principle, be used to distinguish between the two scenarios using sufficiently high-resolution ARPES data
with photon energy dependence used to separate bulk and surface contributions.

We next consider the very interesting quantum oscillation experiments on
Cd$_3$As$_2$ thin films~\cite{analytis:arXiv1505.02817}. The observed Shubnikov-de Haas oscillations seem consistent with ``Weyl orbits'',
which consist of the combination of bulk Landau levels (LLs) and surface Fermi arcs~\cite{Potter2014}.
How can we understand these experimental results, if the surface states have a closed Fermi pocket instead of double Fermi arcs?

First of all, both transport~\cite{Rosenman1969,He2014,analytis:arXiv1505.02817} and ARPES~\cite{Borisenko2014,Neupane:ncomm14} experiments
confirmed that Cd$_3$As$_2$ is slightly n-doped, in the sense that Fermi energy $E_F$
lies above the Dirac points with $k_F\simeq0.04{\AA}^{-1}$. Therefore the surface projection of the bulk electronic states,
near each Dirac point, is in fact a small ``blob'' denoted by a red oval in FIG.~\ref{fig:prl_fig} (d)-(f).
In this case, the surface states at the Fermi energy may (FIG.~\ref{fig:prl_fig}(f)) or
may not (FIG.~\ref{fig:prl_fig}(e)) be connected with the bulk Fermi blobs, depending on $E_F$.
When the surface pocket merges into the bulk blobs, the magnetic orbits responsible for quantum oscillations
must be the Weyl orbits proposed in Ref.~\cite{Potter2014} which consists of both bulk LLs and surface Fermi arcs.
Shubnikov-de Haas oscillations in Ref.~\cite{analytis:arXiv1505.02817} suggest that this is the situation in Cd$_3$As$_2$.
The available ARPES results in Na$_3$Bi~\cite{xu:science15} are also consistent with
surface states merging into bulk projection as $k_z$ increases.

However if we start to introduce p-type dopants to the system and lower the Fermi level, ultimately the surface pocket must
get disconnected from the bulk blobs, as shown in FIG. \ref{fig:prl_fig} (d)-(e), when the Fermi level is close enough to the Dirac points.
In this case the surface Fermi pocket will provide a closed 2D magnetic orbit for cyclotron motion.
This pocket is analogous to the 2D Dirac fermion on the
surface of topological insulators (TIs) and will lead to quantum oscillations similar
to those observed in TIs~\cite{Ren2010,Qu2010,Analytis2010}, with the electron
acquiring a Berry phase of $\pi$ as it goes around the Fermi contour.
As with any 2D Fermi surface, the quantum oscillation frequency $F_s$ has a $1/{\cos\theta}$ dependence
on the angle $\theta$ between magnetic field and surface normal direction.
There is an important difference between the quantum oscillations in the two scenarios
where the bulk and surface states are mixed and separated is that in the latter case, $F_s$
has no dependence on the thickness of the sample, in sharp contrast to the
Weyl orbits~\cite{Potter2014,analytis:arXiv1505.02817} in the former.

Therefore quantum oscillation experiments at different doping levels
provide a sharp distinction between double Fermi arcs and TI-like closed surface Fermi pockets,
serving as an experimental test of our conclusion.
There will be two frequencies in quantum oscillations on DSM thin films~\cite{Potter2014,analytis:arXiv1505.02817}:
$F_b$ associated with bulk Fermi surfaces (due to deviation of $E_F$ from Dirac point) and $F_s$ related to the surface states.
As we move $E_F$ towards bulk Dirac points by doping the system, $F_b$ will always decrease monotonically.
As demonstrated in Ref.~\cite{analytis:arXiv1505.02817}, a triangle-shaped sample does not exhibit
quantum oscillations with frequency $F_s$, while a rectangular sample does, because the
Weyl orbits depend on the thickness of the system along the field direction.
Thus in a triangle-shaped sample, double Fermi arcs cannot lead to quantum oscillation at $F_s$
independent of the Fermi level.
On the other hand, as the surface pocket is disconnected from the bulk blobs by tuning $E_F$ close to the Dirac point
(see FIG. \ref{fig:prl_fig}(e)), the frequency $F_s$, with a 2D angular-dependence, will show up even in a triangular sample.
In particular, when the magnetic length $l_B=\sqrt{\hbar/eB}$ satisfies $k_d\gg l_B^{-1}$ where $k_d$ is the shortest distance between the
surface pocket and the bulk blob in surface BZ, the closed surface pocket cannot tunnel into bulk LLs to form a closed
Weyl orbit. When the Fermi level lies exactly at the Dirac point, $k_d\sim0.04\AA^{-1}$ gives an estimate
of $B<10^2T$ which is within the experimental reach.
Of course, very high-resolution ARPES studies can also directly reveal the separation of
surface Fermi pocket from the bulk blobs as the photon energy is varied.

\section{Concluding remarks}
Let us conclude by summarizing our main results.
Each bulk Dirac node looks like two copies of a Weyl node, which naively suggests that the surface
electronic structure of a DSM should look like two copies of that of the Weyl semimetal (WSM), namely a double Fermi arc.
We show here -- using a simple 4-band model and rigorous K-theory calculations -- that the double
Fermi arcs are not topologically protected, unlike the Fermi arc in a WSM. We find that an arbitrarily small bulk perturbation,
which preserves the Dirac nodes and all the symmetries, can lead to a deformation of the double Fermi arc into a more
conventional Fermi contour. Nevertheless, we show that the surface states in a DSM cannot be completely destroyed,
because there must be topologically protected states on the $k_z=0$ plane, similar to a 3D TI.

Finally, we explore in detail the experimental implications of our results.
We show how we can reconcile our results on the lack of topological protection with existing
experiments on Na$_{3}$Bi~\cite{xu:science15} and Cd$_3$As$_2$~\cite{analytis:arXiv1505.02817}. While these have
been interpreted as being consistent with Fermi arcs, we argue that they are actually a consequence of the mixing of
bulk and surface states.
 We also propose hole-doping the system to disconnect the surface Fermi pocket and bulk states,
and suggest testable signatures for both ARPES and quantum oscillation experiments.\\\\
\textcolor{red}{SI-only references:\cite{Voon2009B,Yu2010B,green_iter}}

\begin{acknowledgments}
MK and MR acknowledge the support of the CEM, an NSF MRSEC, under grant DMR-1420451. YML thanks the Aspen Center for Physics for hospitality, where part of the manuscript was written. YML acknowledges support from startup funds at Ohio State University, and in part from NSF grant PHY-1066293.
\end{acknowledgments}


\end{article}

\end{document}